\documentclass[sn-mathphys,Numbered,iicol]{sn-jnl}

\usepackage{graphicx}%
\usepackage{multirow}%
\usepackage{amsmath,amssymb,amsfonts}%
\usepackage{amsthm}%
\usepackage{mathrsfs}%
\usepackage[title]{appendix}%
\usepackage{xcolor}%
\usepackage{textcomp}%
\usepackage{manyfoot}%
\usepackage{booktabs}%
\usepackage{algorithm}%
\usepackage{algorithmicx}%
\usepackage{algpseudocode}%
\usepackage{listings}%
\usepackage{physics}

\usepackage[T1]{fontenc}
\usepackage[utf8]{inputenc}
\usepackage{placeins,graphicx,dcolumn,bm,xcolor,soul}
\usepackage[caption=false]{subfig}
\usepackage{physics}
\usepackage{appendix}

\begin{document}

\title{Synchronization of phase oscillators due to nonlocal coupling mediated by the slow diffusion of a substance}

\author*[1]{\fnm{Pedro} \sur{Haerter}}\email{haerter@fisica.ufpr.br}

\author[1,2]{\fnm{Ricardo L.} \sur{Viana}}\email{viana@fisica.ufpr.br}

\affil[1]{\orgdiv{Departamento de Física}, \orgname{Universidade Federal do Paraná}, \orgaddress{ \city{Curitiba}, \state{PR}, \country{Brazil}}}
\affil[2]{\orgdiv{Instituto de Física}, \orgname{Universidade de São Paulo}, \orgaddress{ \city{São Paulo}, \state{SP}, \country{Brazil}}}

\abstract{
Many systems of physical and biological interest are characterized by assemblies of phase oscillators whose interaction is mediated by a diffusing chemical. The coupling effect results from the fact that the local concentration of the mediating chemical affects both its production and absorption by each oscillator. Since the chemical diffuses through the medium in which the oscillators are embedded, the coupling among oscillators is non-local: it considers all the oscillators depending on their relative spatial distances. We considered a mathematical model for this coupling, when the diffusion time is arbitrary with respect to the characteristic oscillator periods, yielding a system of coupled nonlinear integro-differential equations which can be solved using Green functions for appropriate boundary conditions. In this paper we show numerical solutions of these equations for three finite domains: a linear one-dimensional interval, a rectangular, and a circular region, with absorbing boundary conditions. From the numerical solutions we investigate phase and frequency synchronization of the oscillators, with respect to changes in the coupling parameters for the three considered geometries. 
}

\maketitle
\section{Introduction}

Phase oscillators are mathematical models of a large number of dynamical systems of physical and biological interest. If a dissipative system, for example, has a stable limit-cycle solution in its phase space, as in the case of self-sustained oscillators, the motion along the limit-cycle can be parameterized by a geometrical phase with a time rate that, in general, depends on the phase itself \cite{1}. The simplest case is of a uniform oscillator, for which the frequency is constant. There are other situations in which we can define such a phase, for example the bursting activity of a neuron \cite{2}. 

There is a wide variety of types of coupling among phase oscillators in a given assembly. If the oscillators occupy fixed positions along a regular lattice, for example, a local type of coupling considering the interaction among nearest-neighbors \cite{3}. In the other extreme, when the interaction among oscillators is practically instantaneous and spatially uniform, each oscillator is coupled to all its neighbors, no matter how far they are. This is the global, or all-to-all coupling \cite{4}. A paradigmatic model of globally coupled phase oscillators was proposed by Kuramoto in 1975, for which a mean-field analytical treatment is possible in the continuum limit \cite{5}. The Kuramoto model has been intensively studied since then, allowing many generalizations and extensions to complex networks \cite{6}.

Key dynamical phenomena in such assemblies of phase oscillators are phase-locking and frequency synchronization, when the uncoupled oscillators have randomly chosen frequencies. In the Kuramoto model, the mean field analysis shows that, in the continuum limit, there is a continuous transition between non-synchronized oscillators and partially synchronized ones, after a critical value of the coupling strength \cite{7}. Similar synchronization transitions were observed for other coupling prescriptions, for complex networks with small-world and scale-free properties \cite{8}. 

Twenty years after his seminal work, Kuramoto proposed a different non-local coupling by considering an assembly of pointlike systems, or cells, each of them undergoing a dynamical evolution with a stable limit-cycle \cite{9}. In this model, the cells occupy fixed positions in the configuration space, and their coupling is mediated through the diffusion of some substance. The latter is both produced and absorbed by each cell, the rates of which depend on their dynamics states \cite{10}. The concentration of the chemical satisfies an inhomogeneous diffusion equation, where the source term takes into account the position of each cell \cite{11}.

This kind of coupling can model a number of systems of biological interest. For example, the suprachiasmatic nucleus (SCN) is a small region of the hypothalamus responsible to synchronize the circadian rhythms with the external $24$ hour light-dark cycle \cite{12}. The SCN is composed by a large number (about ten thousand) of clock cells that adjust their individual rhythms to the photic stimulation \cite{13}. In order to act in a collective way, it is necessary that the clock cells synchronize themselves \cite{14,15,16}. It has been proposed that they do so by releasing and absorbing a neurotransmitter (GABA) which diffuses in the intercell medium \cite{17,18}. 

Another biological example is the observed synchronization of menstrual rhythms displayed by women sharing a common place \cite{19}. One possible explanation for this interesting phenomenon is the release and absorption of pheromones that diffuse through the air, mediating the coupling among menstrual rhythms. However, this issue is still highly controversial in the literature \cite{20,21}. 

One major assumption of the Kuramoto model is that the diffusion is instantaneous, i.e. the diffusion time of the chemical mediating the interaction is much smaller than the characteristic period of the cell dynamics \cite{9}. This assumption is fairly acceptable in the previous example of menstrual rhythm synchronization, which has a circa 28-day period, far greater than the diffusion time, which is of the order of some seconds. This simplifying assumption allows one to eliminate adiabatically the concentration of the mediating chemical and write down a system of non-locally coupled oscillators with some interaction kernel that depends on the Green function of the free boundary value problem \cite{22}. Kuramoto and his collaborators have intensively studied dynamical properties of this nonlocal model for a variety of systems, like nucleation kinetics \cite{23}, multi-scaled turbulence \cite{24}, pattern formation \cite{25}, chemical turbulence \cite{26}, and noise-induced turbulence \cite{27}.

In other situations, however, the diffusion time can be comparable or even larger than the oscillator period. This is particularly interesting when the periods are small, like in the neuronal activity characterized by periodic bursting, for example. In the latter case, one can regard the interaction of neurons as due to chemical synapses, where a neurotransmitter is both released and absorbed in the synaptic cleft \cite{28}. Viana and Aristides have considered the non-local coupling mediated by the diffusion of a substance when the diffusion time is finite, and with absorbing boundary conditions for a number of geometries in a configuration space, with specified initial profiles \cite{29}. 

The diffusion equation is solved by using an appropriate Green function for finite domains with Dirichlet boundary conditions. There resulted a system of coupled nonlinear integro-differential equations, which takes into account all the history of the phase evolution for each time. In the infinite-time limit our results reduce to those obtained by Nakao in the fast-relaxation (adiabatic) case \cite{22}.  

The system of coupled integro-differential equations introduced in Ref. \cite{29} was approximately solved by an iterative procedure, in which the evolution of the uncoupled phase oscillators is taken as a zeroth-order solution. The latter is plugged back into the coupling integral, yielding analytical results for the first-order solution, obtained from a system of coupled differential equation (Born approximation) \cite{29}. In the present work, we revisit this treatment with an improved numerical scheme for directly solving the integro-differential equations. In addition, we considered also different geometries, including two-dimensional domains with rectangular and circular boundaries.

For each geometry used, the solution of the integro-differential equation is used to investigate numerically both phase and frequency synchronization properties of the network. We compute the order parameter magnitude of the network, and analyze its dependence with the coupling parameters of the model, namely the coupling strength, the diffusion coefficient, and the degradation parameter. Using this methodology the system presents a loss in synchronization when either the degradation or diffusion parameters increase their values, and also the enhancement of the synchronization effect with the increase of the coupling strength.

\begin{figure*}
	\centering
	\includegraphics[scale=0.7]{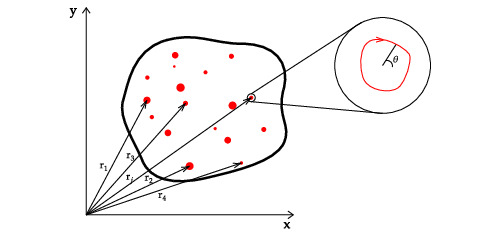}
	\caption{Schematic figure of an assembly of pointwise phase oscillators randomly distributed over a bounded spatial domain.}
	\label{fig:diagrama_osc}
\end{figure*}

The rest of this paper is organized as follows: in Section II we outline the coupling model, showing the integro-differential equation to be solved for an assembly of phase oscillators in one and two spatial dimensions, and within finite domains with absorbing boundary conditions. Section III is devoted to a discussion of phase and frequency synchronization properties with respect to the coupling parameters. Our Conclusions are left to last Section.

\section{Integro-differential equations for nonlocal coupling}
\label{sec:Modelo}

Let us consider a system which presents a stable limit-cycle in its phase space, like a self-sustained dissipative oscillator  \cite{nakaonovo}. For such systems, a phase $\theta \in [0,2\pi)$ can be defined as a geometrical angle parameterizing motion on the limit cycle. There are, however, other ways to define a phase that can applied to other situations, e.g. the bursting of a neuron \cite{nossoantigo}. Although the frequency $\omega$, as the time rate of the phase, can be a function of $\theta$, we will consider them as constants in a lowest order approximation. 

The mathematical model we deal with in the present work consists of $N$ pointlike phase oscillators, embedded in a region of a $d$-dimensional Euclidean space, with positions ${\bf r}_j$, where $j=1, 2, \ldots N$ [Fig. \ref{fig:diagrama_osc}]. Each oscillator is described by a phase $\theta_j$ whose evolution is given by ${\dot\theta}_j = \omega_j$, where $\omega_j$ is a frequency randomly chosen from a given probability distribution function $g(\omega)$. We assume that $g(-\omega) = g(\omega)$ with a maximum at $\omega = 0$ \cite{5}.  

We suppose that the phase oscillators both produce and absorb a given chemical, in such a way that the rates of production and absorption both depend on the oscillator frequencies. The coupling among oscillators depends on the local concentration of this substance, which is a scalar field $A({\bf r},t)$, resulting in the following system of differential equations 
\begin{equation}
	\frac{d\theta_j}{dt} = \omega_j+ K A({\bf r}_j,t) \qquad (j = 1, 2, \ldots N),
	\label{eq:EDO_X_A}
\end{equation}
where $K > 0$ is a coupling strength, measuring the influence of the local concentration of the diffusing chemical in the dynamics of each phase oscillator. We assume that the concentration satisfies an inhomogeneous  diffusion equation of the form \cite{11}
\begin{equation}
	\frac{\partial A}{\partial t}+ \eta A - D \nabla^2 A = \sum^{N}_{k=1} h(\theta_k) \delta({\bf r} - {\bf r}_k),
	  \label{diffusion}
\end{equation}
where $D > 0$ is the diffusion coefficient, $\eta > 0$ is a degradation coefficient, and there is a source term representing the effect of the pointwise oscillators, each of them producing the chemical with an intensity  given by a function $h$ of their phases \cite{22}. 

In order to solve this equation in a bounded spatial domain ${\cal R}$ we specify appropriate boundary conditions at some limiting surface $\partial{\cal R}$, as well as an initial condition profile $A({\bf r},t=0)$. The Green function for Eq. (\ref{diffusion}), $G({\bf r},t;{\bf r'},t')$, satisfies \cite{duffy}
\begin{equation}
  \label{green}
  \frac{\partial G}{\partial t} + \eta  \, G - D \,  \nabla^2 G = \delta({\bf r}-{\bf r'})  \, \delta(t-t'),
\end{equation}
for homogeneous Dirichlet boundary conditions: $G({\bf r},t;{\bf r'},t')$ for ${\bf r} \in {\partial{\cal R}}$, and the initial condition $G({\bf r},t=0; {\bf r'},t') = 0$. The solution of the inhomogeneous diffusion equation (\ref{diffusion}) within the domain ${\cal R}$, for absorbing boundary conditions in $\partial{\cal R}$, $A({\bf r}\in\partial{\cal R},t) = 0$, and an initial profile $A({\bf r},t=0)=0$, is given by \cite{29}
\begin{equation}
	A({\bf r},t)=\sum^N_{k=1} \int^{t}_{0} dt' \, G({\bf r},t|{\bf r}_k,t') \, h(\theta_k(t')). 
\end{equation}

Substituting this solution into Eq. (\ref{eq:EDO_X_A}) there results a system of integro-differential equations governing the coupling among phase oscillators mediated by the diffusion of a substance \cite{29}
\begin{equation}
	\frac{d\theta_j}{dt} = \omega_j + K \sum^N_{k=1} \int^{t}_{0} dt' G({\bf r}_j,t;{\bf r}_k,t') \, h(\theta_k(t')),
	\label{eq:integro}
\end{equation}
where $j = 1, 2, \ldots N$. In this work we choose a source function which makes contact with the classical Kuramoto model \cite{5}
\begin{equation}
 \label{choiceH}
 h(\theta_k) = \frac{1}{N}  \sin(\theta_k - \theta_j),
\end{equation} 
for which (\ref{eq:integro}) reads $(j = 1, 2, \ldots N)$
\begin{equation}
  \label{phase1}
  \frac{d\theta_j}{dt} = \omega_j + \frac{K}{N} \sum_{k=1}^N \int_0^{t} dt' \sin[\theta_k(t') - \theta_j(t')] G({\bf r}_j,t;{\bf r}_k,t').
\end{equation}

In the following we consider three different geometries for the limited domain in which the phase oscillators are embedded. 

{\bf 1. One-dimensional domain:} the Dirichlet Green function for a one-dimensional finite domain $0 \le x \le L$ with absorbing boundary conditions ($A(0,t)=A(L,t)=0$) is \cite{duffy}
\begin{align}
    \nonumber     
    G(x,t;x',t') & = \frac{2H(t-t')}{L} \sum_{n=1}^\infty \sin\left(\frac{n\pi x}{L}\right) \sin\left(\frac{n\pi x'}{L}\right) \times \\
    \label{fini1}
    & \exp\left\{ - \left\lbrack D{\left(\frac{n\pi}{L}\right)}^2 + \eta\right\rbrack (t-t') \right\},
\end{align}
where $H(t-t')$ is the Heaviside unit-step function. In our numerical simulations we consider an assembly of $N$ oscillators, occupying randomly chosen positions ${\{ x_j\}}_{j=1}^N$ within the interval $0 < x < L$. The coupled system of equations is given by (\ref{phase1}) as 
\begin{equation}
  \label{phase1a}
  \frac{d\theta_j}{dt} = \omega_j + \frac{K}{N} \sum_{k=1}^N \int_0^{t} dt' \sin[\theta_k(t') - \theta_j(t')] G(x_j,t;x_k,t'),
\end{equation}
where $j = 1, 2, \ldots N$ and the Green function is given by (\ref{fini1}). The natural frequencies $\omega_j$ are randomly chosen according to a Gaussian  distribution with zero mean and variance $0.02$. The initial phases $\theta_i(0)$ (with $i=1, 2, \ldots N$) are also randomly chosen according to a uniform probability distribution in the interval $[0,2\pi)$. 

{\bf 2. Two-dimensional rectangular domain:} given by $0 < x < a$, $0 \le y \le b$. The corresponding Dirichlet Green function with absorbing boundary conditions is \cite{duffy} 
\begin{align}
 \nonumber
 G({\bf r},t;{\bf r'},t') = \frac{4H(t-t')}{L} \sum_{n=1}^\infty \sum_{m=1}^\infty \sin\left(\frac{n\pi x}{a}\right) 
 \times \\
 \label{fini2}
 \sin\left(\frac{n\pi x'}{a}\right) \sin\left(\frac{m\pi y}{b}\right) \sin\left(\frac{m\pi y'}{b}\right) \times \\
 \nonumber
 \exp\left\{ - \left\lbrack D\left(\frac{n^2}{a^2}+\frac{m^2}{b^2}\right)\pi^2 + \eta\right\rbrack (t-t') \right\}.
\end{align}
The resulting system of equation is similar to (\ref{phase1a}), in which the phase oscillators have randomly chosen positions ${\{ x_j,y_j\}}_{j=1}^N$ within the rectangle $0 < x < a$, $0 \le y \le b$. The integro-differential equations (\ref{phase1}) are complemented by the Green function (\ref{fini2}). The initial phases and frequencies are chosen in a similar way, and the numerical procedure is essentially the same as before. 

{\bf 3. Two-dimensional circular domain:} a circle of radius $r = a$, with the Dirichlet Green function (with absorbing boundary conditions) given by (in polar coordinates) \cite{duffy} 
\begin{align}
 \nonumber
 G(r,\vartheta,t;r',\vartheta',t') = \frac{1}{\pi D} \sum_{m=-\infty}^\infty \sum_{n=1}^\infty \frac{1}
 {{[{J'}_m(x_{mn})]}^2} \times \\ 
 \label{greencirc1}
 J_m\left(x_{mn} \frac{r}{a} \right) J_m\left(x_{mn} \frac{r'}{a} \right)
 \cos[m(\vartheta-\vartheta')] \times \\
 \nonumber 
 \exp\left\{ -\left(\eta+\frac{D x_{mn}^2}{a^2} \right) \,  (t-t') \right\},
 \end{align}
where $x_{mn}$ is the $n$th positive root of the Bessel function $J_m$. The randomly chosen positions, in polar coordinates, are ${\{ r_j,\vartheta_j\}}_{j=1}^N$ such that $0 < r < a$ and $0 < \theta < 2\pi$, and enter the Green function given by (\ref{greencirc1}). 

In the fast-diffusion limit \cite{30}, for which the concentration of the coupling substance attain its equilibrium value instantaneously, the expressions we obtained for the Green functions reduce to those proposed by Kuramoto and Nakao \cite{10,11,22}. 

\section{Phase and frequency synchronization}
\label{sec:Res}

\begin{figure*}[!t]
  \centering{\includegraphics[scale=0.40]{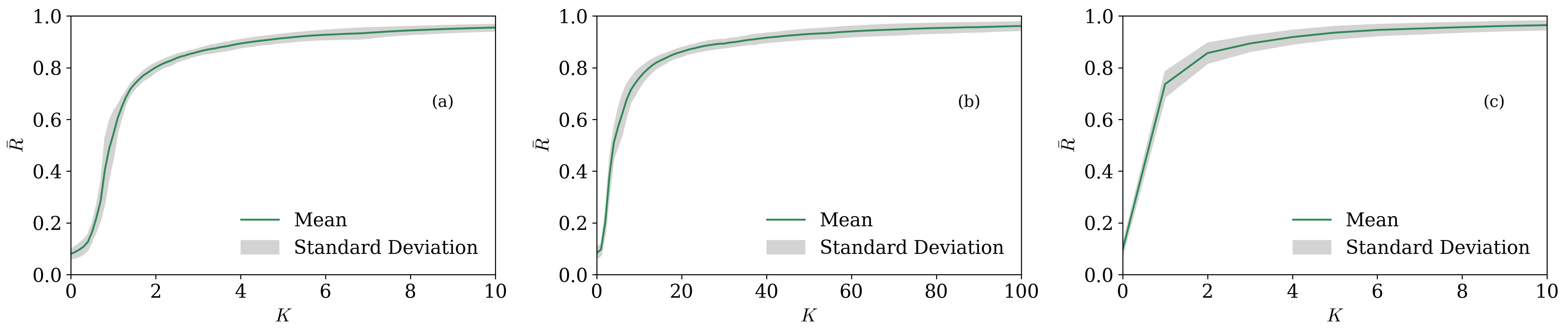}}
  \caption{Mean order parameter magnitude {\it vs.} coupling strength for assemblies of $N = 100$ phase oscillators randomly distributed over (a) linear, (b) rectangular, and (c) circular domains, with absorbing boundary conditions} \label{2}
\end{figure*}

One of the more important collective phenomena displayed by assemblies of coupled phase oscillators is synchronization of their phases and frequencies. If each oscillator is regarded as a kind of physical or biological clock, their assemblies are expected to exhibit coherent oscillations in order to generate a common rhythm. This is the case, for example, in the suprachiasmatic nucleus (SCN): each clock cell must synchronize its individual rhythms in order to produce a coherent output. 

One numerical diagnostic of phase synchronization is provided by the Kuramoto complex order parameter\cite{6}, defined as 
\begin{equation}
 \label{orderp}
 z(t) = R(t) \, e^{i\psi(t)} = \frac{1}{N} \sum_{k=1}^N e^{i\theta_k(t)},
\end{equation}
where $\theta_k(t)$ is the phase of the $k$th oscillator at time $t$. The order parameter can be regarded as the vector sum of phasors, with magnitude $R(t)$ and angle $\psi(t)$. Since, for finite $N$, the order parameter magnitude exhibits size-dependent fluctuations, we compute its  temporal mean 
\begin{equation}
 \label{orderpm}
 {\bar R} = \lim_{T\rightarrow\infty} \frac{1}{T-T'} \int_{T'}^T R(t) dt,
\end{equation}
where $T'$ is the duration of a transient period. If ${\bar R} \approx 0$ the oscillators are completely non-synchronized in phase, representing phasors uniformly distributed along the interval $[0,2\pi)$ for each time $T' < t < T$. On the other hand, if ${\bar R} \approx 1$ the phasors add coherently and the oscillators are completely phase-synchronized. Intermediate values of ${\bar R}$ characterize partial phase synchronization. 

Figure \ref{2} shows the variation of the mean order parameter magnitude ${\overline R}$ with the coupling parameters, for assemblies of $N = 100$ phase oscillators with diffusion and degradation coefficients $D = \eta = 1$, respectively. The curves represent an average over different realizations of the randomly chosen initial conditions for the phases, whereas the sidebands stand for the respective standard deviations. For a one-dimensional finite domain with $L = 1$ we numerically solved Eq. (\ref{phase1a}) with the Green function given by (\ref{fini1}). The order parameter magnitude exhibits a monotonic increase of ${\overline R}$ with the coupling strength $K$, signaling a synchronization transition roughly at $K = 1$ [Fig. \ref{2}(a)]. 

The $K\rightarrow 0$ limit of the order parameter is nonzero for all domains: this is due to a number of spurious synchronizations occurring by chance. This is a typical effect of considering systems with small number $N$ of individuals, such that for $N\rightarrow\infty$ we expect to have ${\overline R} \approx 0$ when $K \rightarrow 0$.  As $K$ increases the phase oscillators become progressively synchronized and the order parameter magnitude tends to the unity. The same trend is also observed for the rectangular domain [of sides $a = b = 1$], but the range of $K$ is ten times higher than for the linear domain [Fig. \ref{2}(b)]. This is clearly a consequence of the increasing spatial size, in the sense that in two-dimensions it takes a higher value of $K$ to achieve the same synchronization effect than for the linear case. However, this effect seems not to be valid for a circular domain [Fig. \ref{2}(c)], for which the $K$-range is similar to the linear case. Apparently this is due to the radial symmetry of the circular domain, turning it more akin to the one-dimensional case than the rectangular domain.

\begin{figure*}[!t]
  \centering{\includegraphics[scale=0.40]{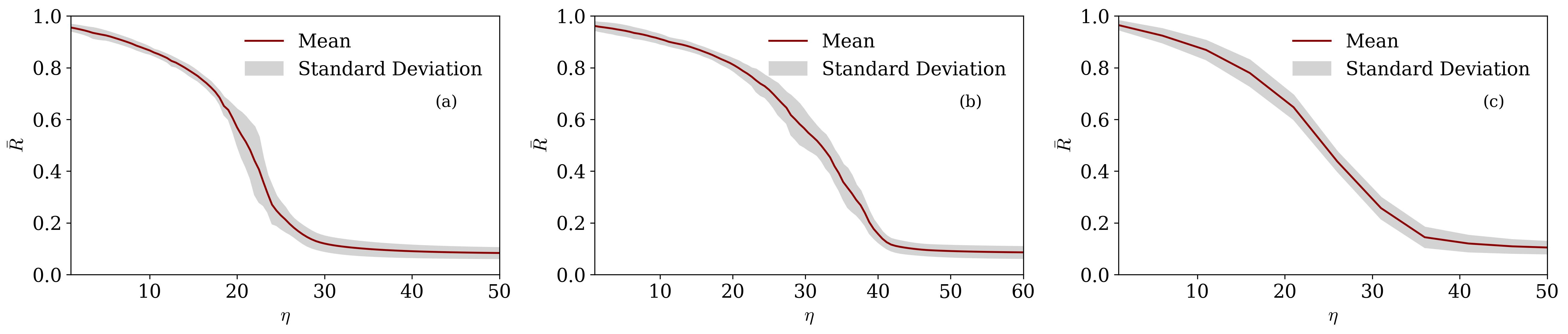}}
  \caption{Mean order parameter magnitude {\it vs.} degradation parameter for assemblies of $N = 100$ phase oscillators randomly distributed over (a) linear, (b) rectangular, and (c) circular domains, with absorbing boundary conditions}\label{3}
\end{figure*}

The values of ${\overline R}$ are plotted In Figure \ref{3}, as a function of the degradation coefficient $\eta$, for fixed values of $K = 10$ and $D = 1$. In all domains we observe a decrease in the mean order parameter magnitude as $\eta$ increases. Since the degradation coefficient measures the loss of the substance mediating the coupling, the basic effect of its increase is the decrease in the amount of phase synchronization. The range of $\eta$ for the three domains is nearly the same for the linear [Fig. \ref{3}(a)] and circular domains [Fig. \ref{3}(c)], with a slight difference for the rectangular domain [Fig. \ref{3}(b)], for which the desynchronization occurs later than in the other cases. We see in these results another example of the already mentioned similarity between linear and circular domains.

\begin{figure*}[!t]
  \centering{\includegraphics[scale=0.40]{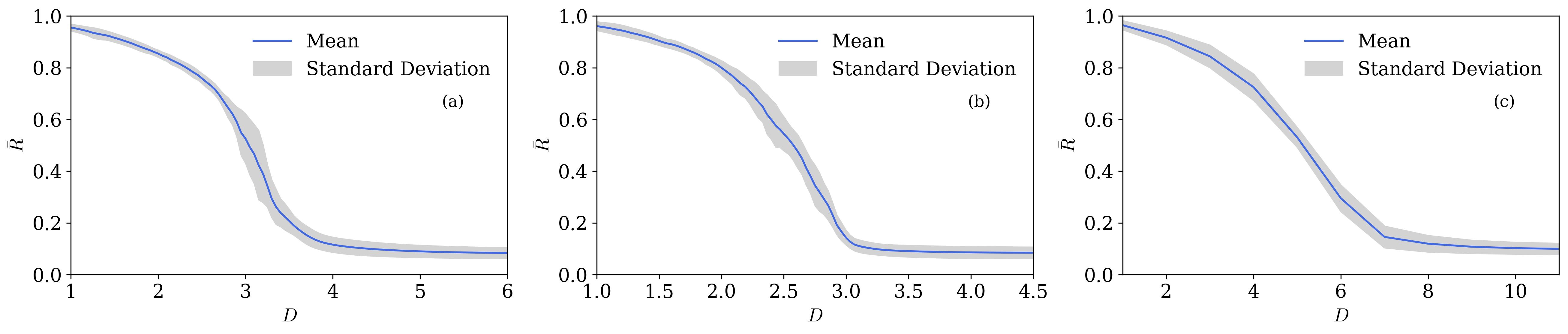}}
  \caption{Mean order parameter magnitude {\it vs.} diffusion coefficient for assemblies of $N = 100$ phase oscillators randomly distributed over (a) linear, (b) rectangular, and (c) circular domains, with absorbing boundary conditions.} \label{4}
\end{figure*}

We plot, in Fig. \ref{4}, the values of ${\overline R}$ as a function of the diffusion coefficient $D$, for fixed $K = 10$ and $\eta = 1.0$. We started at $D = 1.0$ in order to avoid a singular behavior characteristic of the limit $D \rightarrow 0$. Just like in the previous figure, the values of ${\overline R}$ decrease monotonically as $D$ increases, i.e. the effect of larger $D$ is to actually  desynchronize the oscillator phases. This can be qualitatively explained by considering that the coupling effect is more effective the longer the mediating substance remains in the spatial medium in which the oscillators are embedded. There are three factors influencing this permanence time: firstly, the degradation of the mediating substance decreases its concentration with time, so difficulting synchronization. In the second place, the absorbing boundary conditions act as a sink, removing any amount of the mediating substance that reaches the domain boundary. The third factor is the diffusion coefficient $D$, since if the latter is large, the permanence time of the substance is smaller, and thus reduces the synchronization effect as well. 

These considerations can be made quantitative by computing the characteristic diffusion time $\tau_D$. From the diffusion equation (\ref{diffusion}) without sources, the latter can be estimated (for a one-dimensional domain of length $L$) as 
\begin{equation}
 \label{diftime}
 \tau_D = {\left( \frac{D}{L^2} - \eta \right)}^{-1},
\end{equation} 
Hence if $D$ increases, for fixed $\eta$, the diffusion time decreases, leading to a shorter permanence of the coupling substance in the domain (thanks to the degradation and absorbing boundaries) and thus to a poor synchronization effect. This explains why the order parameter magnitude actually decreases with $D$. Such result has been also observed in our previous numerical simulations made with the Born approximation.

\begin{figure*}
	\centering{\includegraphics[scale=0.40]{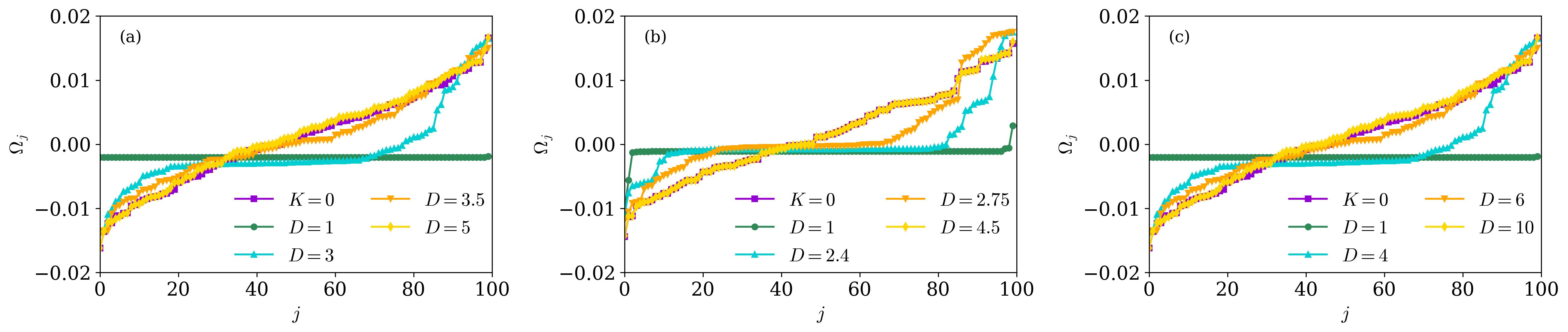}}
  \caption{Perturbed frequencies of $N = 100$ phase oscillators for different values of the diffusion coefficient in  (a) linear, (b) rectangular, and (c) circular domains, with absorbing boundary conditions.} \label{5}
\end{figure*}

\begin{figure*}
	\centering{\includegraphics[scale=0.40]{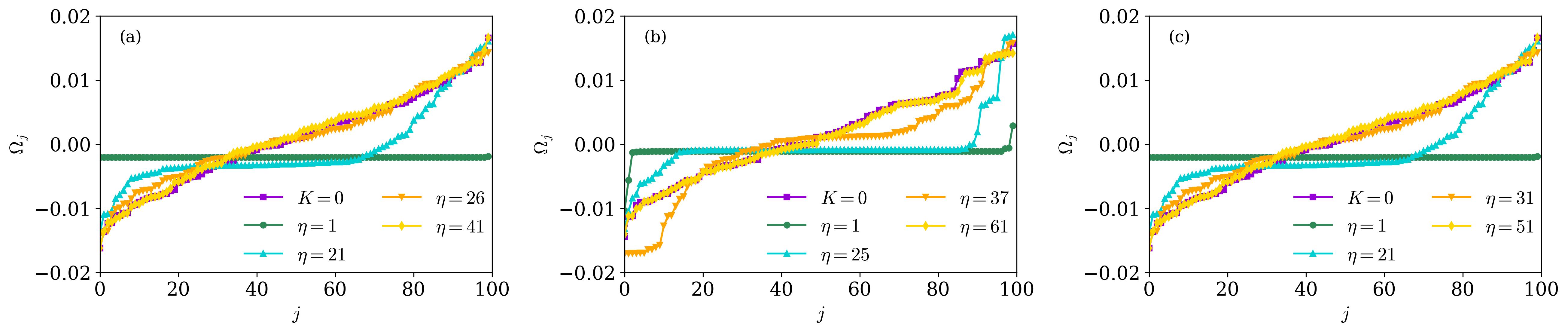}}
  \caption{Perturbed frequencies of $N = 100$ phase oscillators for different values of the degradation coefficient in  (a) linear, (b) rectangular, and (c) circular domains, with absorbing boundary conditions.} \label{6}
\end{figure*}

Finally the effect of the domain geometry on the desynchronization that occurs when the diffusion coefficient increases is less pronounced than for the previous case. The intervals of $D$ which lead to a steeper decrease in ${\overline R}$ are: (i) from $3$ to $4$ for the linear case [Fig. \ref{4}(a)]; (ii) from $2$ to $3$ for the rectangular domain  [Fig. \ref{4}(b)]; (iii) from $4$ to $6$ for the circular geometry [Fig. \ref{4}(c)]. 

The frequencies of the coupled oscillators can be computed directly from the definition
\begin{equation}
\label{omegas}
 \Omega_j = \lim_{t\rightarrow\infty} \frac{1}{t} \left\{  \theta_j(t+T')-\theta_j(t) \right\}, \qquad (j=1, 2, \ldots N),
\end{equation}
where $T'$ is such that transients have decayed. For zero coupling it turns out that $\Omega_j(t) = \omega_j$, whereas a finite coupling makes a number of oscillator to equalize their frequencies around the zero mean, forming a plateau of zero-frequency oscillators. Phase synchronization implies frequency synchronization, but the converse is not necessarily true (the phases can be locked at fixed values, for example, with a constant phase difference). We have computed numerically the perturbed frequencies from Eq. (\ref{omegas}) and shown that, for the three spatial domains the results are in accordance with those exhibited for the order parameter, which measures phase synchronization. 

In Fig. \ref{5} and Fig. \ref{6} we plot the oscillator frequencies in increasing order of their values. The case of vanishing coupling $(K=0)$ is shown for the sake of comparison. The small asymmetry observed in this case is an effect of the finite size of the oscillator assembly used in numerical simulations. Figure \ref{5}(a) displays the values of $\Omega_j$ for $K = \eta = 1.0$ and different values of the diffusion coefficient for the linear domain. We see that large values of $D$ yield results similar to the uncoupled case, confirming that the effect of increasing $D$ is to desynchronize the oscillators [see Fig. \ref{4}(a)]. The same trend is present in the results for the rectangular and circular domains, displayed in Figs. \ref{5}(b) and \ref{5}(c), respectively. The same behavior can be seen in Fig. \ref{6} for the degradation coefficient. 

\section{Conclusions}
\label{sec:con}

We considered in this work some aspects of the spatio-temporal dynamics of pointlike phase oscillators coupled by a diffusible substance. Such coupling occurs in the manner that the oscillators will produce and absorb an amount of a substance depending on their dynamics. In previous works, it has been considered that the diffusion would be so fast (as compared with the characteristic periods of the oscillators) that the concentration achieves immediately its equilibrium value, yielding a non-local type of coupling which depends on the relative distances among oscillators. 

In the present work, however, we take into account a finite diffusion time, which demands solving simultaneously the oscillator equations coupled with the diffusion equation. This produces a system of nonlinear integro-differential equations. We solved these equations using a full numerical code that takes into account the entire history of the concentration evolution in order to update the oscillators phases and frequencies, in the case of a nonlinear coupling similar to that used in the classical Kuramoto model. 

Among the various collective phenomena displayed in the course of the spatio-temporal evolution of the network, we choose to investigate numerically the synchronization of phases and frequencies in the assembly of coupled phase oscillators. This choice is motivated by the example of the clock cells belonging to the suprachiasmatic nucleus in the mammal brain, that have their own individual rhythms but need to synchronize them so as to produce a coherent output. In particular, we numerically investigated the role of three coupling parameters on the synchronization properties of the network: the coupling strength, and the diffusion and degradation coefficients. The numerical diagnostics used were the order parameter magnitude and the dispersion of the perturbed oscillator frequencies. 

We also considered three different geometries: a one-dimensional finite interval (linear) and two-dimensional bounded domains (rectangular and circular), where we used absorbing boundary conditions. This choice was chiefly motivated by the relative simplicity of the Green functions involved in the model formulation, which have been written as a rapidly converging eigenfunction series expansion. If we consider reflecting boundary conditions, the Green functions would have to be substantially modified, and even in the one-dimensional case, their form would be so complicated that different numerical methods would be preferable, like finite differences. 

The dependence of the order parameter magnitude with the coupling strength is similar to that displayed by the classical Kuramoto model. For weak coupling the oscillators remain non-synchronized and start to exhibit partial phase synchronization after a critical coupling value. The precise determination of the latter would need the use of a large number of oscillators, which is presently unfeasible due to limitations in computer time. However, it can be estimated that the critical coupling strength in the two-dimensional rectangular case is about ten times higher than for the one-dimensional linear and the two-dimensional circular case. 

Moreover, the order parameter magnitude decays with the degradation coefficient for all three geometries. This is due to the desynchronizing effect of a loss of the coupling substance due to its degradation into other, non-active chemicals. A similar decay of the order parameter was observed when the diffusion coefficient is increased. This seemingly contradictory result is a consequence of the absorbing boundary conditions: a rapidly diffusing substance will disappear once it reaches the boundaries, so their staying time would decrease and thus the corresponding effect on the improvement of synchronization among oscillators. If we were to consider reflecting (or partially absorbing) boundary conditions, the increase of the concentratin within the spatial domain would eventually lead to a saturation, and so an enhanced synchronization. 

Although the present work has focused on the synchronization of phase oscillators, our model of diffusion-mediated coupling can be extended for virtually any dynamical system, both in continuous or discrete time. Using, for example, nonlinear equations in problems of chemical kinetics would lead to models of chemical turbulence involving the local diffusion of substances. Such numerical investigations could, for example, focus on spatio-temporal patterns and a wide number of instability phenomena. 

\section*{Acknowledgements}

One of the authors (R. L. V.) would like to thank Professor I. L. Caldas for useful discussions and suggestions. P. H. thanks CAPES for partial financial support. R. L. V. received partial financial support from the following Brazilian government agencies: CNPq (403120/2021-7, 301019/2019-3), CAPES (88881.143103/2017-01), and FAPESP (2022/04251-7).

\appendix*
\section{Numerical Method}
The integro-differential equations for the coupled phase oscillators has the general form
\begin{equation}
    \dv{\theta_j}{t}=f_j(t,\theta_j(t))+\int_{0}^{t}F_j(t,s,\theta_j(s))\dd s \qquad (j = 1, 2, \ldots N),
\label{eq:integro_generico} 
\end{equation}
making each time point an infinitesimal position of size $h$, so that $\theta_j(t_n)$ is the value of the function $\theta_j$ at time $t_n=nh$, just as $\theta^{'}_j(t_n)$ is the derivative of $\theta$ at time $t_n$ and $\theta_j^n$ represents the approximation of $\theta_j$ at time $t_n$. Integrating the equation between two infinitesimal times $t_n$ and $t_n+h$

\begin{align}
\begin{split}
\theta_j(t_n+h)&=\theta_j(t_n)+\mathcal{I}_j^{(1)}+\mathcal{I}_j^{(2)}
 \end{split}
 \end{align} 
where

\begin{align}
    \mathcal{I}_j^{(1}=\int_{t_n}^{t_n+h}f_j(t,\theta_j(t))\dd t\\
    \mathcal{I}_j^{(2)}=\int_{t_n}^{t_n+h}\int_{0}^{t}F_j(t,s,\theta_j(s))\dd s\dd t
\end{align}

For the $t$ integrals we integrated using the trapezoidal method, 

\begin{equation}
    \mathcal{I}_j^{(1)} = \frac{h}{2}\qty[f_j(t_n,\theta_j(t_n))+f_j(t_{n+1},\theta_j(t_{n+1}))]
\end{equation}
and

\begin{align}
\begin{split}
    \mathcal{I}_j^{(2)} = \frac{h}{2} \Bigg[\int_{0}^{t_n}F_j(t_n,s,\theta_j(s))&\dd s+\\
    \int_{0}^{t_{n+1}}F_j&(t_{n+1},s,\theta_j(s))\dd s\Bigg]
    \end{split}
\end{align}
however, the value $\theta_j(t_{n+1})$ is unknown, so we use the formal definition of derivative, obtaining the approximation $\theta_{j}^{n+1}=\theta_j^{n}+\theta_j^{'n}$, and using a first approximation of the equation (\ref{eq:integro_generico})

\begin{equation}
    \theta_j^{'n+1}=f_j(t_{n+1},\theta_j^{n+1})+\int_{0}^{t_{n+1}}F_j(t_{n+1},s,\theta_j(s))\dd s
    \label{eq:integral_j}
\end{equation}

The integral in (\ref{eq:integral_j}) can be solved numerically by using the trapezoidal method again. Applying its result in $\mathcal{I}_j^{(2)}$ and using $\theta_j^{n+1}=\theta_j^{n}+h\theta_j^{'n}$ the final equation for $\theta_j(t_{n}+h)$ will be

\begin{align}
    \begin{split}
        \theta_j&(t_n+h)=\theta_j(t_n)+\frac{h}{2}\qty[f_j(t_n,\theta_j^n)+f_j(t_{_n+1},\theta_j^{n}+h\theta_j^{'n})]\\
        &+\frac{h^2}{4}\qty[F_j(t_n,0,\theta_j^0)+\ldots+F_j(t_n,t_n,\theta_j^n)]\\
    &+\frac{h^2}{4}\qty[F_j(t_{n+1},0,\theta_j^0)+\ldots+F_j(t_{n+1},t_{n+1},\theta_j^{n}+h\theta_j^{'n})]
    \end{split}
    \label{eq:completo_int}
 \end{align} 

 By choosing the appropriate value of $h$ it is possible to have results of considerable precision and closer to reality, the main point being to balance between precision and execution time.

\end{document}